\documentclass[aps,floatfix,showpacs,nofootinbib,,amsmath,amssymb]
{revtex4}
\usepackage[dvips]{graphicx}

\begin{document}

\title{Light-cone averages in a swiss-cheese universe}

\author{Valerio Marra} \email{valerio.marra@pd.infn.it}

\affiliation{Dipartimento di Fisica ``G.\ Galilei'' Universit\`{a} di Padova,
INFN Sezione di Padova, via Marzolo 8, Padova I-35131, Italy}
\affiliation{Department of Astronomy and Astrophysics, the 
University of Chicago, Chicago, IL \ \ 60637-1433 }

\author{Edward W.\ Kolb} \email{rocky.kolb@uchicago.edu} 
\affiliation{ Department of Astronomy and Astrophysics, Enrico Fermi
Institute, and  Kavli Institute for Cosmological Physics, the University of
Chicago, Chicago, IL \ \ 60637-1433 }

\author{Sabino Matarrese} \email{sabino.matarrese@pd.infn.it}
\affiliation{Dipartimento di Fisica ``G.\ Galilei'' Universit\`{a} di Padova,
INFN Sezione di Padova, via Marzolo 8, Padova I-35131, Italy}

\begin{abstract}

We analyze a toy swiss-cheese cosmological model to study the averaging
problem.  In our swiss-cheese model, the cheese is a spatially flat, matter
only, Friedmann-Robertson-Walker solution (\textit{i.e.,} the Einstein--de
Sitter model), and the holes are constructed from a Lema\^{\i}tre-Tolman-Bondi
solution of Einstein's equations. We study the propagation of photons in the
swiss-cheese model, and find a phenomenological homogeneous model to describe
observables. Following a fitting procedure based on light-cone averages, we
find that the the expansion scalar is unaffected by the inhomogeneities
(\textit{i.e.,} the phenomenological homogeneous model is the cheese model). 
This is because of the spherical symmetry of the model; it is unclear whether
the expansion scalar will be affected by non-spherical voids.  However, the
light-cone average of the density as a function of redshift is affected by 
inhomogeneities.  The effect arises because, as the universe evolves,  a photon
spends more and more time in the (large) voids than in the (thin) high-density
structures.  The phenomenological homogeneous model describing the light-cone
average of the density is similar to the $\Lambda$CDM concordance model.  It is
interesting that although the sole source in the swiss-cheese model is matter,
the phenomenological homogeneous model behaves as if it has a dark-energy
component.  Finally, we study how the equation of state of the phenomenological
homogeneous model depends on the size of the inhomogeneities, and find that 
the equation-of-state parameters $w_{0}$ and $w_{a}$ follow a power-law
dependence with a scaling exponent equal to unity. That is, the equation of
state depends linearly on the distance the photon travels through voids. We
conclude that within our toy model, the holes must have a present size of about
$250$ Mpc to be able to mimic the concordance model. 

\end{abstract}


\maketitle

\section{Introduction}

Most, if not all, observations are consistent with the cosmic concordance model,
according to which  one-fourth of the present mass-energy of the universe is
clustered and dominated by cold dark matter (CDM).  The remaining
three-quarters is uniform and dominated by a fluid with negative pressure (dark
energy, or $\Lambda$). 

While the standard $\Lambda$CDM model seems capable of accounting for the
observations, 95\% of the mass-energy of the present universe is unknown. This
is either a feature, and we are  presented with the opportunity of discovering
the nature of dark matter and dark energy, or it is a bug, and nature might be
different than described by the $\Lambda$CDM model. Regardless, until such time
as dark matter and dark energy are completely understood, it is useful to look
for alternative cosmological models that fit the data.

One non-standard possibility is that there are large effects on the {\it
observed} expansion rate (and hence on other observables) due to the
back-reaction of inhomogeneities in the universe (see, \textit{e.g.,} Ref.\
\cite{kmr,notari,rasa, buchert} and references therein).  The basic idea is that
all evidence for dark energy comes from the observational determinations of the
expansion history of the universe.  Anything that affects the observed expansion
history of the universe alters the determination of the parameters of dark
energy; in the extreme it may remove the need for dark energy.

The ``safe'' consequence of the success of the concordance model is that the
isotropic and homogeneous $\Lambda$CDM model is a good  {\it phenomenological}
fit to the real inhomogeneous universe. And this is, in some sense, a
verification of the cosmological principle:  the inhomogeneous universe can be
described by means of an isotropic and  homogeneous solution.  However, this
does not imply that a primary source of dark energy  exists, but only that it
exists as far as the  phenomenological fit is concerned. For example, it is not
straightforward that the universe is accelerating. If dark energy does not
exist at a fundamental level, its presence in the concordance model would tell
us that the pure-matter inhomogeneous model has been renormalized, from the
phenomenological point of view  (luminosity-distance and redshift of photons),
into a homogeneous $\Lambda$CDM model.

The issue is the observational significance of the back-reaction of
inhomogeneities. Our point of view is tied to our past light cone: we focus on
the effects  of large-scale nonlinear inhomogeneities on observables such as
the  luminosity-distance--redshift relation. We will not discuss averaged
domain dynamics, even though if we think it is a crucial step in understanding
how General Relativity effectively works in a  lumpy universe \cite{ellis,
buchert_new}.

Following this approach, we built in Ref.\ \cite{marra-sc} a particular
swiss-cheese model, where the cheese consists of a spatially flat, matter only
Friedmann-Robertson-Walker (FRW) solution and the holes are constructed out of
a Lema\^{i}tre-Tolman-Bondi  (LTB) solution of Einstein's equations. We
attempted to find a model that was solvable and ``realistic'' (even if still
toy), rather than finding a model with interesting volume-averaged
dynamics. The model, however, will turn out to be useful to investigate
light-cone averages.

It has been indeed shown that the LTB solution can be used to fit the observed
$d_{L}(z)$  without the need of dark energy (for example, see Ref.\
\cite{alnes0602}).  To achieve this result, however, it is necessary to place
the observer at the center of a rather large-scale underdensity. To overcome
this  fine-tuning problem we built a swiss-cheese model with the observer in
the cheese looking through a series of holes.

In Ref.\ \cite{marra-sc} we studied this model in detail and discussed the
effects of large-scale nonlinear inhomogeneities on observables such as the
luminosity-distance--redshift relation.  We found that inhomogeneities are able
(at least partly) to mimic the effects of dark energy.

In this paper we will analyze the same swiss-cheese model through the fitting
scheme developed by Ellis and Stoeger \cite{ellis-f} in order to better
understand how inhomogeneities renormalize the (matter only) swiss-cheese
model allowing us to avoid a physical dark-energy component.  We think that
this model fits well in that context and therefore we  might be able to shed
some light on the important topics discussed there. We will propose a fitting
procedure based on light-cone averages.

The paper is organized as follows: In Sec.\ \ref{model} we will specify the
parameters of our swiss-cheese model and summarize the main results  obtained
in Ref.\ \cite{marra-sc}. In Sec.\ \ref{fitti}, we develop our fitting
procedure, and in  Sec.\ \ref{disco} we discuss our results. Then, in Sec.\
\ref{dressing} we study the dependence of the best-fit  parameters on the size
of the holes. Conclusions are given in Sec.\ \ref{conclusions}.

\section{The swiss-cheese model} \label{model}

In this section we will briefly describe the model studied in Ref.\ 
\cite{marra-sc}; we refer the reader there for a more thorough analysis. In our
swiss-cheese model, the cheese consists of a spatially flat, matter only, 
Friedmann--Robertson--Walker solution, and the spherically symmetric holes are
constructed from a Lema\^{\i}tre-Tolman-Bondi solution.

In Table \ref{units} we list the units we will use for mass density, time, the
radial coordinate, the expansion rate, and two quantities, $Y(r,t)$ and $W(r)$,
that will appear in the metric. The time $t$ appearing in Table \ref{units} is
not the usual time in FRW models.  Rather, $t=T-T_0$, where $T$ \textit{is} the
usual cosmological time and $T_0=2H_0^{-1}/3$ is the present age of the
universe.  Thus, $t=0$ is the present time and $t=t_{BB}=-T_0$ is the time of
the big bang.  Finally, the initial time of the LTB evolution is defined as
$\bar{t}$. 

Both the FRW and the LTB metrics can be written in the form (in the synchronous
and comoving gauge)
\begin{equation}
ds^2 = -dt^{2}+\frac{Y'^2(r,t)}{W^2(r)}dr^2+Y^2(r,t) \, d\Omega^2 ,
\end{equation}
where here and throughout, the ``prime'' superscript denotes $d/dr$ and the 
``dot'' superscript will denote $d/dt$.  It is clear that the Robertson--Walker
metric is recovered with the substitution $Y(r,t)=a(t)r$ and $W^2(r)=1-kr^2$.

\begin{table}
\caption{\label{units} Units for various quantities.  We use 
geometrical units, $c=G=1$. Here, the present critical density is
$\rho_{C0}=3H^{2}_{0,\, Obs}/8 \pi$, with $H_{0,\, Obs}=70 \textrm{ km
s}^{-1}\textrm{ Mpc}^{-1}$. In order to have the proper distance today 
we have to multiply the comoving distance by $a(t_{0})\simeq 2.92$.}
\begin{ruledtabular}
\begin{tabular}{lccr}
Quantity          & Notation    & Unit            & Value            \\ 
\hline
mass density & $\rho(r,t)$, $\bar{\rho}(r,t)$ & $\rho_{C 0}$ 
& $9.2\times10^{-30}\textrm{ g cm}^{-3}$             \\
time              & $t$, $T$, $\bar{t}$, $t_{BB}$, $T_0$ 
& $(6 \pi \rho_{C 0})^{-1/2}$  & $9.3\textrm{ Gyr}$ \\
comoving radial coordinate & $r$         & $(6 \pi \rho_{C 0})^{-1/2}$  
& $2857 \textrm{ Mpc}$ \\
metric quantity   & $Y(r,t)$    & $(6 \pi \rho_{C 0})^{-1/2}$  
& $2857 \textrm{ Mpc}$ \\
expansion rate    & $H(r,t)$    & $(6 \pi \rho_{C 0})^{1/2} $ 
& $\frac{3}{2}H_{0,\, Obs}$ \\
spatial curvature term    & $W(r)$      & $1$     &        ---             \\
\end{tabular}
\end{ruledtabular}
\end{table}

\subsection{The cheese}

We choose for the cheese model a spatially-flat,  matter-dominated universe
(the Einstein--de Sitter (EdS) model).   In the cheese there is no $r$
dependence to $\rho$ or $H$. Furthermore, $Y(r,t)$ factors into a function of
$t$ multiplying $r$ ($Y(r,t) = a(t)r$), and in the EdS model $W(r)=1$.  In this
model $\Omega_{M}=1$, so in the cheese, the value of $\rho$ today, denoted as
$\rho_0$, is unity in the units of Table \ref{units}.   In order to connect
with the LTB solution, we can express the line element in the form
\begin{equation}
ds^2=-dt^2+Y'^2(r,t) dr^2 + Y^2(r,t) \, d\Omega^2 .
\end{equation}

In the cheese, the Friedmann equation and its solution are
\begin{eqnarray}
H^{2}(t) & = & \frac{4}{9} \; \rho(t)=\frac{4}{9}(t+1)^{-2} \\
Y(r,t)& = & r \, a(t)=r \, \frac{(t+1)^{2/3}}{(\bar{t}+1)^{2/3}},
\end{eqnarray}
where the scale factor is normalized so that at the beginning of the LTB
evolution it is $a(\bar{t})=1$.

\subsection{The holes}

\subsubsection{The General LTB model}

The holes are chosen to have a LTB metric \cite{lemaitre, tolman, bondi}.  
The model is based on the assumptions that the system is  spherically symmetric
with purely radial motion and the motion is geodesic without shell crossing
(otherwise we could not neglect the pressure).

It is useful to define an ``Euclidean'' mass $M(r)$ and an ``average'' mass
density $\bar\rho(r,t)$, as
\begin{equation}
M(r) = 4\pi \int_0^r \rho(r,t) \: Y^2 Y' \: dr
= \frac{4 \pi}{3} Y^{3}(r,t) \: \bar{\rho}(r,t) .
\end{equation}
In spherically symmetric models, in general there are two expansion rates: an
angular expansion rate, $H_\perp\equiv \dot{Y}(r,t)/Y(r,t)$, and a radial
expansion rate, $H_r\equiv \dot{Y}'(r,t)/Y'(r,t)$.  (Of course in the FRW model
$H_r=H_\perp$.)
The angular expansion rate is given by
\begin{equation}
H^2_\perp(r,t) = \frac{4}{9} \; \bar{\rho}(r,t) +\frac{W^2(r)-1}{Y^{2}(r,t)}\;.
\label{motion}
\end{equation}
Unless specified otherwise, we will identify $H_\perp=H$.

To specify the model we have to specify initial conditions, \textit{i.e.,} the
position $Y(r,\bar{t})$, the velocity $\dot{Y}(r,\bar{t})$ and the density
$\rho(\bar{t})$ of each shell $r$ at time $\bar{t}$. In the absence of  shell
crossing it is possible  to give the initial conditions at different times for
different shells $r$: let us call this time $\bar{t}(r)$. The initial conditions
fix the arbitrary curvature function $W(r)$:
\begin{equation}
\label{cucu}
W^2(r)-1 \equiv 2 E(r)= \left. \left(\dot{Y}^2-  
\frac{1}{3 \pi}\frac{M}{Y}\right)\right|_{r,\bar{t}} \ ,
\end{equation}
where we can choose $Y(r,\bar{t})=r$ so that $M(r) = 4 \pi
\int_{0}^{r}\rho(\bar{r},\bar{t}) \: \bar{r}^{2} \: d\bar{r}$.

In a general LTB model there are therefore three arbitrary 
functions\footnote{One of these three functions only expresses the gauge 
freedom as discussed in Ref.\ \cite{marra-sc}, Appendix A.}:
$\rho(r,\bar{t})$, $W(r)$ and $\bar{t}(r)$. Their values for the 
particular LTB model we study are specified in the following subsection.

\subsubsection{Our LTB model} \label{ourmodel}

First of all, for simplicity we choose $\bar{t}(r)=\bar{t}$; \textit{i.e.,} we
specify the initial conditions for each shell at the same moment of time.

We now choose $\rho(r,\bar{t})$ and $W(r)$ in order to match the flat FRW model
at the boundary of the hole: \textit{i.e.,} at the boundary of the hole
$\bar{\rho}$ has to match the FRW density and $W(r)$ has to go to unity.  A
physical picture is that, given a FRW sphere, all the matter in the inner
region is pushed to the border of the sphere while the quantity of matter
inside the sphere does not change. With the density chosen in this way, an
observer outside the hole will not feel the presence of the hole as far as
\textit{local} physics is concerned (this does not apply to global quantities,
such as the luminosity-distance--redshift relation). In this way we can imagine
putting in the cheese as many holes as we want, even with different sizes and
density profiles,  and still have an exact solution of the Einstein equations
(as long as there is no superposition among the holes and the correct matching
is achieved). So the cheese is evolving as an FRW universe while the holes
evolve differently.  This idea was first proposed by Einstein and Straus
\cite{einstein}.

As anticipated in the Introduction we are building in this way a model exactly
solvable and ``realistic''  (even if still toy) at the price of not having any
interesting volume-averaged  dynamics. The volume evolution of this
swiss-cheese model is indeed unaffected  by the inhomogeneities. We are not
concerned about this because we think that average dynamics  is not {\it
directly} correlated to observable quantities. We will see however that this
model will be interesting for light-cone averages.

\begin{figure}
\begin{center}
\includegraphics[width=12cm]{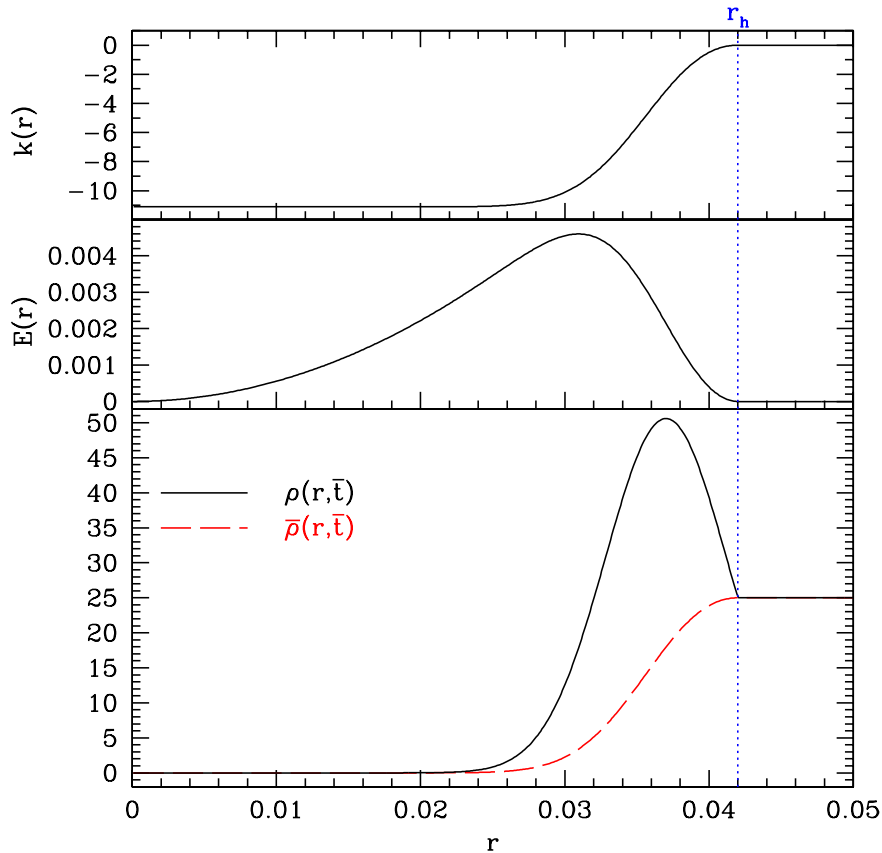}
\caption{Bottom: The densities $\rho(r,\bar{t})$ (solid curve) and
$\bar{\rho}(r,\bar{t})$  (dashed curve). Here, $\bar{t}=-0.8$ (recall
$t_{BB}=-1$). The hole ends at $r_{h}=0.042$. The matching to the FRW solution
is achieved  as one can see from the plot of $\bar{\rho}(r,\bar{t})$.
Top: Curvature $k(r)$ and $E(r)$ necessary for the 
initial conditions of no peculiar velocities.}
\label{rhoek}
\end{center}
\end{figure}

\begin{figure}
\begin{center}
\includegraphics[width=15cm]{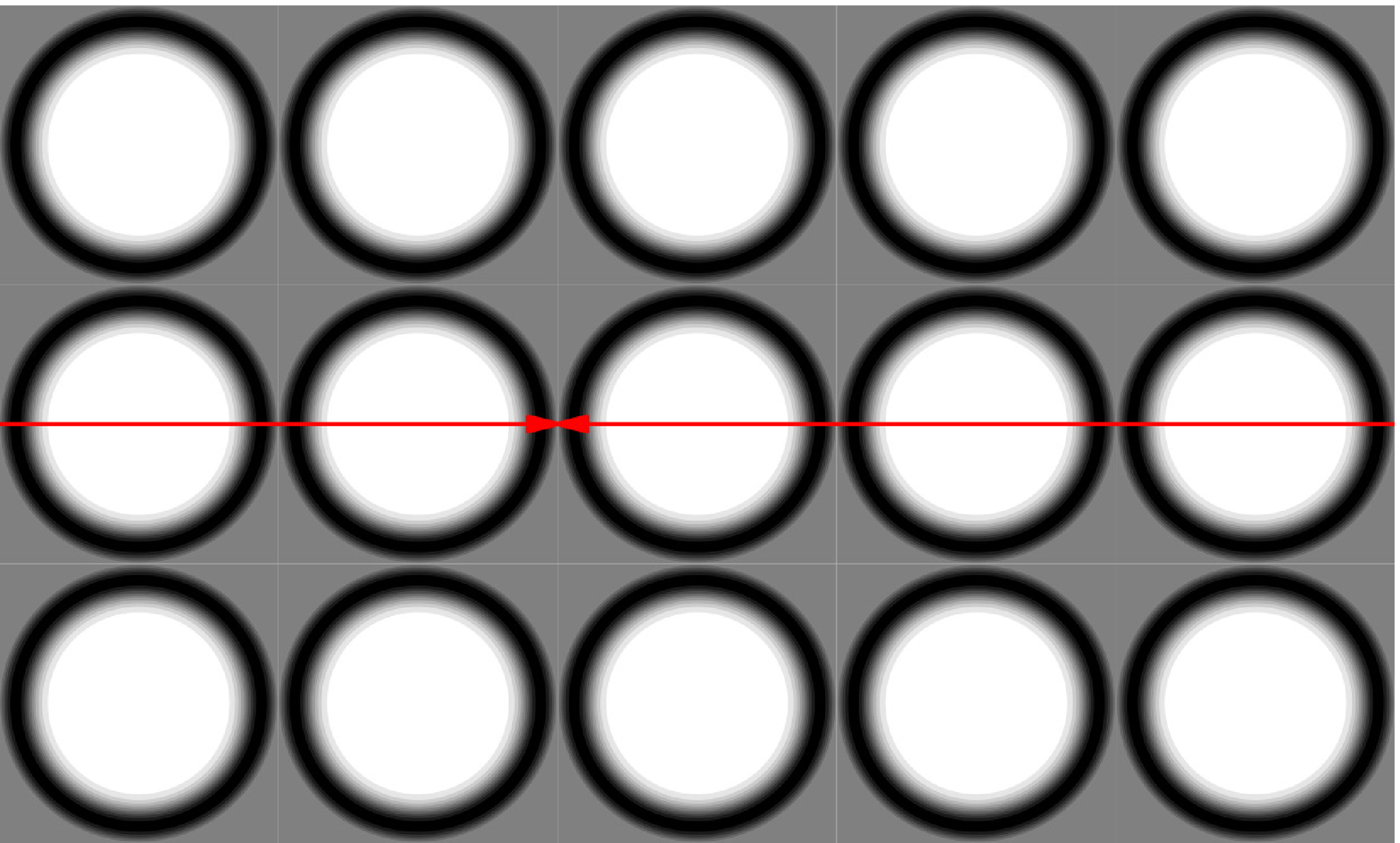}
\caption{Sketch of our model. The shading mimics 
the initial density profile: darker shading implies larger denser. The uniform 
gray is the FRW cheese. The photons pass through the holes as shown by the 
arrows and are revealed by the observer placed in the cheese.}
\label{schizzo}
\end{center}
\end{figure}

In Fig.\ \ref{rhoek} we plot the chosen Gaussian density profile. The hole ends
at $r_{h}=0.042$, which is $350$ Mpc in size,\footnote{To get this number from
Table \ref{units}, multiply $r_{h}$ by $a(t_{0})\simeq 2.92$.} roughly $25$
times smaller than $r_{BB}$. Note that this is not a very big bubble. But it is
an almost empty region: in the interior the matter density is roughly $10^4$
times smaller than in the cheese. Our model consists of a sequence of five
holes with the observer looking through them. The idea, however, is that the
universe is completely filled with  these holes, which form a sort of lattice.
In this way an observer in  the cheese will see an isotropic  CMB along the two
directions of sight shown in Fig.\ \ref{schizzo}.

To have a realistic evolution, we demand that there are no initial peculiar
velocities at time $\bar{t}$, \textit{i.e.,} that the initial expansion $H$
is independent of $r$. This implies
\begin{equation} 
\label{Er}
E(r)=\frac{1}{2} H_{FRW}^2(\bar{t}) \, r^{2}-\frac{1}{6\pi}\frac{M(r)}{r}.
\end{equation}
The function $E(r)$ chosen in this way is shown in Fig.\ \ref{rhoek}. 
As seen from the figure, the curvature $E(r)$ is small compared with unity. 
In spite of its smallness, the curvature plays a crucial role to
allow a realistic evolution of structures.

In Fig.\ \ref{rhoek} we also plot $k(r)=-2 E(r)/r^{2}$, which is the
generalization of the factor $k$ in the usual FRW models (it is not normalized
to unity). As one can see, $k(r)$ is very nearly constant  in the empty region
inside the hole. This is another way to see the reason for our choice of the
curvature function: we want to have in the center an empty bubble dominated by
negative curvature.

\subsection{The dynamics} \label{dynamics}

\begin{figure}
\begin{center}
\includegraphics[width=16.2 cm]{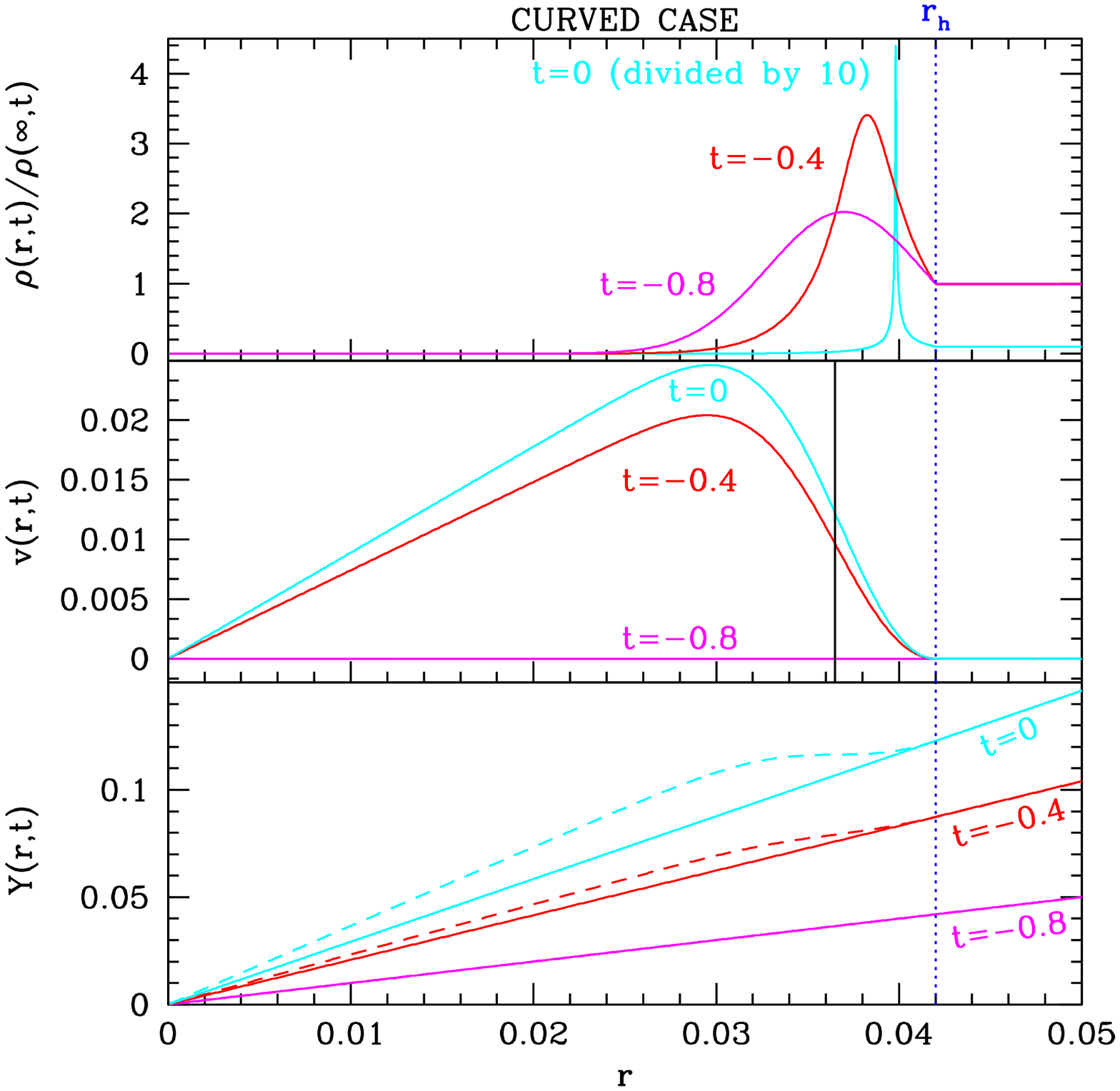}
\caption{Behavior of $Y(r,t)$ with respect to $r$,  the peculiar velocities
$v(r,t)$ with respect to $r$, and the density profiles $\rho(r,t)$ with respect
to $r_{FRW}=Y(r,t)/a(t)$, for the curved case at times $t=\bar{t}=-0.8$, 
$t=-0.4$ and $t=t_0=0$. The straight lines for $Y(r,t)$ are the FRW solutions
while the dashed lines are the LTB solutions.  For the peculiar velocities, the
matter gradually starts to move toward high density regions. The solid vertical
line marks the position of the peak in the density with respect  to $r$. For the
densities, note that the curve for $\rho(r,0)$ has been divided by $10$.  
Finally, the values of $\rho(\infty,t)$ are $1,\ 2.8,$ and $25$, for $t=0,\
-0.4,\ -0.8$,  respectively. }
\label{curved}
\end{center}
\end{figure}

In Fig.\ \ref{curved} we show the evolution of $Y(r,t)$ for three times: 
$t=\bar{t}=-0.8$ (the Big Bang is at $t_{BB}=-1$), $t = -0.4$, and $t=0$ 
(corresponding to today). From Fig.\ \ref{curved} it is clear that outside the
hole, \textit{i.e.,} for $r \geq r_{h}$, $Y(r,t)$ evolves as a FRW solution,
$Y(r,t)\propto r$.

The inner almost empty region is expanding faster than the outer (cheese)
region. The density ratio between the cheese and the interior region of the hole
increases by a factor of $2$ between $t=\bar{t}$ and $t=0$. Initially the
density ratio was $10^{4}$, but the model is not sensitive to this number since
the evolution in the interior region is dominated by the curvature ($k(r)$ is
much larger than the matter density).

The evolution is realistic, as one can see from Fig.\ \ref{curved},  which
shows the density profile at different times. Overdense regions start
contracting and become thin shells (mimicking structures), while underdense
regions become larger (mimicking voids), and eventually occupy most of the
volume.

Let us explain why the high density shell forms and the nature of shell
crossing. Because of the distribution of matter, the inner part of the hole is
expanding faster than the cheese; between these two regions  there is the
initial overdensity. It is because of this that there is less matter in the
interior part. (Remember that we matched the FRW density at the end of the
hole.) Now we clearly see what is happening: the overdense region is squeezed
by the interior and exterior regions, which act as a clamp. Shell crossing
eventually happens when more shells---each labeled by its own $r$---are so
squeezed that they occupy the same physical position $Y$, that is when $Y'=0$.
Nothing happens to the photons other than passing through more shells at the
same time: this is the meaning of the $g_{r r}$ metric coefficient going to
zero.

Remember that $r$ is only a label for the shell whose Euclidean position at time
$t$ is $Y(r,t)$.  In the plots of the energy density we have normalized $Y(r,t)$
using $r_{FRW}=Y(r,t)/a(t)$.

\subsection{Redshift histories} \label{histories}

\begin{figure}
\begin{center}
\includegraphics[width=12 cm]{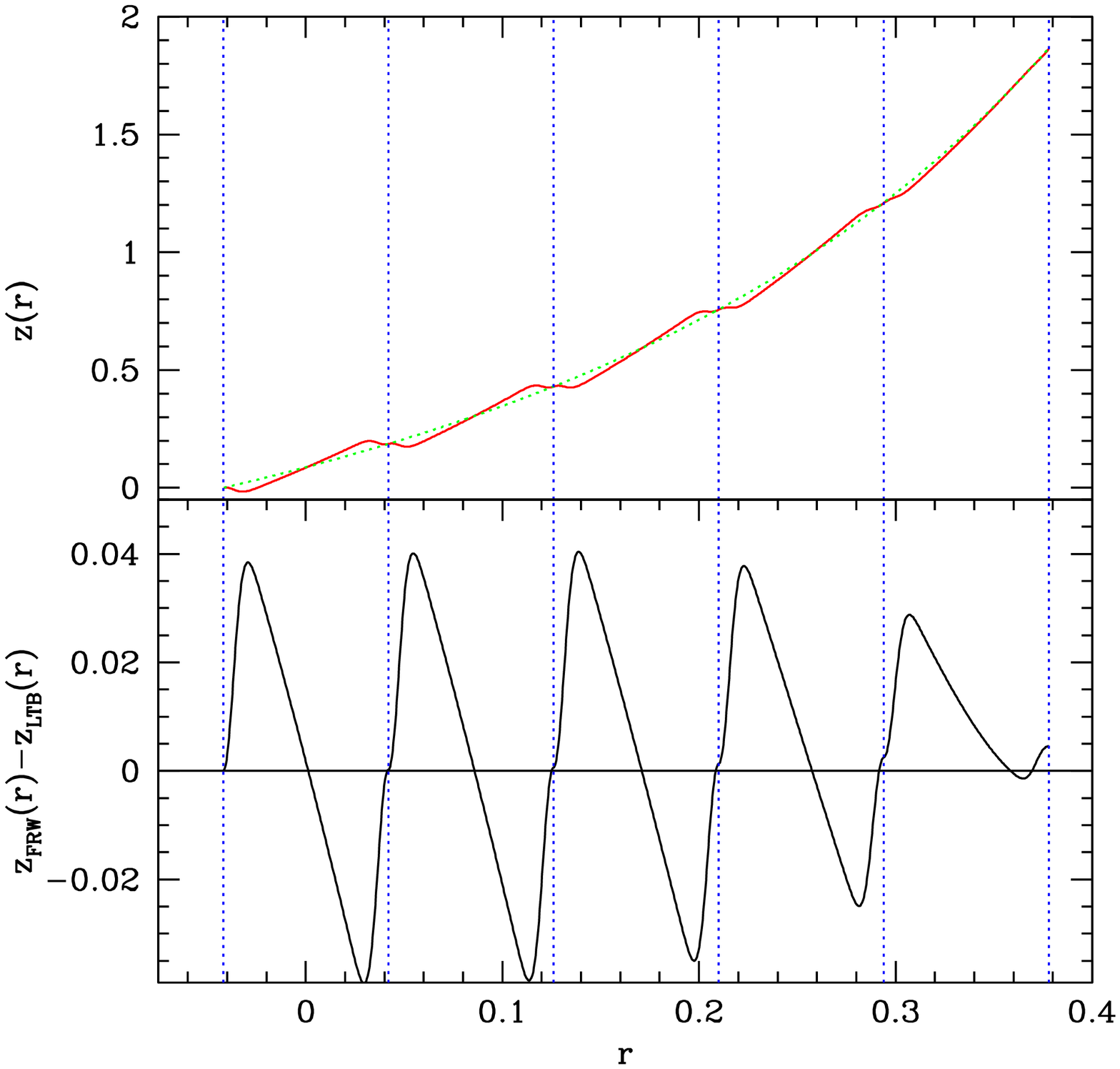}
\caption{Redshift histories for a photon that travels from one 
side of the five-hole chain to the other where the 
observer will detect it at present time. The dotted curve is for the FRW
model. The vertical lines mark the edges of the holes. The plots are with
respect to the coordinate radius $r$. Notice also that along the voids the
redshift is increasing  faster: indeed $z'(r)=H(z)$ and the voids are expanding
faster.}
\label{zorro}
\end{center}
\end{figure}

As shown in Fig.\ \ref{zorro}, this model does not feature substantial redshift
effects: it is anyhow natural to expect a compensation, due to the spherical
symmetry, between the incoming path and the outgoing path inside the same hole.

However, there is a compensation already on the scale of half a hole as it is
clear from the plots. This mechanism is due to the density profile chosen, that
is one whose average matches the FRW density of the cheese: roughly speaking we
know that $z'=H \propto \rho = \rho_{\scriptscriptstyle FRW} + \delta \rho$. We
chose the density profile in order to have $\langle \delta \rho \rangle=0$, and
therefore in its journey from the center to the border of the hole the photon
will see a $\langle H\rangle \sim H_{\scriptscriptstyle FRW}$ and therefore
there will be compensation for $z'$.

Let us see this analytically. We are interested in computing a line average of
the expansion along the photon path in order to track what is going on.
Therefore, we shall not use the complete expansion scalar:
\begin{equation}
\theta=\Gamma_{0k}^{k}=2\frac{\dot{Y}}{Y}+\frac{\dot{Y}'}{Y'} ,
\end{equation}
but, instead, only the part of it pertinent to a radial line average:
\begin{equation} \label{dito}
\theta_r=\Gamma_{01}^{1}=\frac{\dot{Y}'}{Y'}\equiv H_{r} ,
\end{equation}
where $\Gamma_{0k}^{k}$ are the Christoffel symbols and $\theta$ is the trace
of the extrinsic curvature.

Using $H_r$, we obtain:
\begin{equation}
\langle H_r \rangle =
\frac{\int_{0}^{r_{h}}dr \; H_r \; Y' / W}
{\int_{0}^{r_{h}} dr \; Y' / W} \simeq \left. 
\frac{\dot{Y}}{Y}\right|_{r=r_{h}}
= H_{\scriptscriptstyle FRW} ,
\end{equation}
where the approximation comes from neglecting the (small) curvature and the last
equality holds thanks to the density profile chosen. This is exactly the result
we wanted to find. However, we have performed an average  at constant time and
therefore we did not let the hole and its structures  evolve while the photon is
passing; the evolution of structures will partially break  this compensation.

We have, therefore, seen that the compensation in redshift on the scale of half
a hole is due to the density profile chosen.  Even if we relax this
requirement, we will still have a compensation on the scale of the hole. This
can be seen in Fig.\ \ref{zorro}: inside each hole, the plot is anti-symmetric
with respect to the center of the hole (the center of symmetry). This is only
approximate at early times when structure evolution is fast enough to change
the second half of the hole with respect to the first half.

This discussion sheds light on the fact that photon physics seems to be affected
by the evolution of the inhomogeneities more than by the inhomogeneities
themselves. We can argue that there should be perfect compensation if the hole
will have a static metric such as the Schwarzschild one. In the end, this is a
limitation of our assumption of spherical symmetry.

\subsection{Luminosity and Angular-Diameter Distances}

\begin{figure}
\begin{center}
\includegraphics[width=16.2 cm]{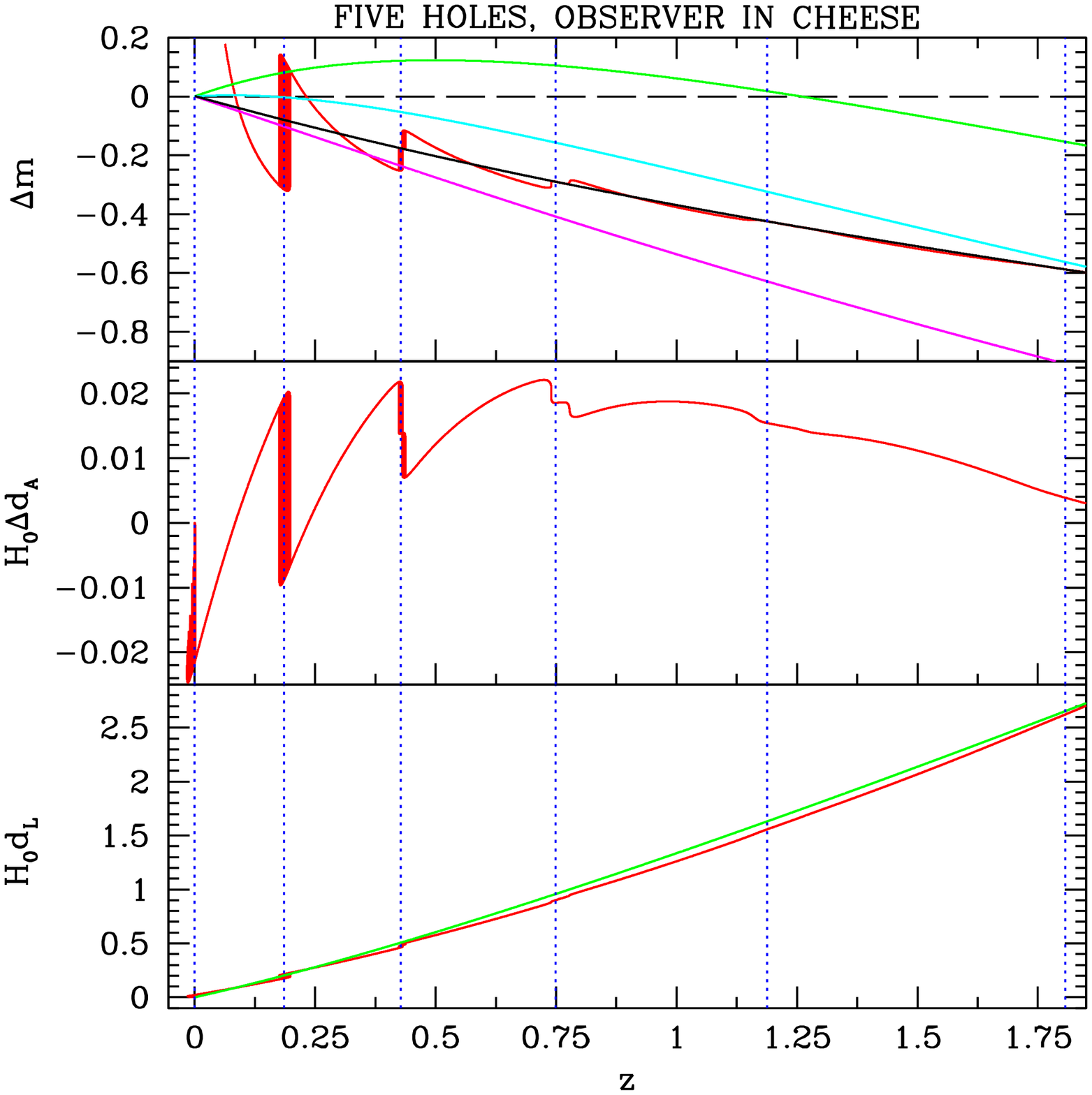}
\caption{On the bottom the luminosity distance $d_L(z)$ in the  five-hole model
(jagged curve) and the $\Lambda$CDM solution with $\Omega_{M}=0.6$ and
$\Omega_{DE}=0.4$ (regular curve) are shown.  In the middle is the change in
the angular diameter distance, $\Delta d_A(z)$, compared to a $\Lambda$CDM
model with  $\Omega_{M}=0.6$ and $\Omega_{DE}=0.4$. The top panel shows the
distance modulus in various cosmological models. The jagged line is for the
five-hole LTB model. The regular curves, from top to bottom, are a $\Lambda$CDM
model with  $\Omega_{M}=0.3$ and $\Omega_{DE}=0.7$, a $\Lambda$CDM model with  
$\Omega_{M}=0.6$ and $\Omega_{DE}=0.4$, the best smooth fit to the LTB model,
and the EdS model.  The vertical lines mark the edges of the five holes.}
\label{5incheese}
\end{center}
\end{figure}

We show in Fig.\ \ref{5incheese} the results for the luminosity distance and
angular distance. The solution is compared to the one of the $\Lambda$CDM model
with $\Omega_{M}=0.6$ and $\Omega_{DE}=0.4$. It has an effective
$q_{0}=\Omega_{M}/2-\Omega_{DE}=-0.1$.

The distance modulus is plotted in the top panel of Fig.\ \ref{5incheese}. The
solution shows an oscillating behavior that is due to the simplification of
this toy model in which all the voids are inside the holes and all the
structures are in thin spherical shells. For this reason a fitting curve was
plotted: it is passing through the points of the photon path that are in the
cheese between the holes. Indeed, they are points of average behavior and
represent well the coarse graining of this oscillating curve. The
simplification of this model tells us also that the most interesting part of
the plot is farthest from the observer, let us say at $z>1$. In this region we
can see the effect of the holes clearly: they move the curve from the EdS
solution to the $\Lambda$CDM one with $\Omega_{M}=0.6$ and $\Omega_{DE}=0.4$.
Of course, the model in not realistic enough to reach the ``concordance''
solution.

Summarizing, because of our assumption of spherical symmetry,  we found no
significant  redshift effects. The effects we found came from the
angular-diameter  distance which is affected by the evolution of the
inhomogeneities.

\section{The fitting problem} \label{fitti}

Now that we have seen how the luminosity-distance--redshift relation is 
affected by inhomogeneities, we want to study the same model from the  point of
view of light-cone averaging to see if we can gain insights into how 
inhomogeneities renormalize the matter swiss-cheese model and mimic a
dark-energy component.

As explained in Ref.\ \cite{ellis-f}, there are, broadly speaking, two
distinct approaches that have been applied to understand the large-scale
structure of the universe.

The standard approach is to make the assumption of spatial homogeneity and
isotropy on a large enough scale, and to assume this guarantees that that the
universe is represented by a FRW model. In other words, it is assumed that the
dynamics of an inhomogeneous universe with density $\rho(\vec{x})$ is identical
to the dynamics of a homogeneous universe of density
$\langle\rho(\vec{x})\rangle$. The main problem with this approach is that it
simplifies the way the real lumpy universe should be averaged. It does not
really specify any type of averaging procedure necessary to make use of the FRW
model, and it assumes that, in any case, the dynamics is not affected by
inhomogeneities.  Therefore, there is no information on what scales such a
model is supposed to be applicable, if any.

The concordance model fits very well the experimental data: the direct
consequence of its success is, indeed, that the isotropic and homogeneous
$\Lambda$CDM model is a good {\it phenomenological}  fit to the real
inhomogeneous universe. And this is, in some sense, a reflection of the
cosmological principle of spatial homogeneity and isotropy on a large enough
scale:  the inhomogeneous universe can be described by means of a isotropic and
homogeneous solution. However this does not imply that a primary dark energy
component really  exists, but only  that it exists effectively as far as the
phenomenological fit is concerned. For example, it is not an observational
consequence that the universe is globally  accelerating (although it is usually
stated as such).  If primary dark energy does not exist, observational
evidence coming from the concordance model would tell us that the pure-matter
inhomogeneous model  has been renormalized from the phenomenological point of
view (\textit{e.g.,} the luminosity-distance and redshift of photons), into a
homogeneous $\Lambda$CDM model. 

The other approach is to make no \textit{a priori} assumption of global
symmetry, and build up our universe model only on the basis of astronomical 
observations. The main problem with the such an approach is the practical 
difficulty in implementing it. 

An approach which is intermediate between the two outlined  above is based on
the fitting procedure.  It asks the question about which FRW  model
best fits our lumpy universe. This question will  lead to a procedure
that  will allow us to understand better how to interpret the large-scale FRW
solution.

The best-fit procedure will be implemented along the past light cone. This is
because a meaningful fitting procedure should be related directly to
astronomical observations. 

A remark is in order here: in the previous section we did not fit 
the $d_L(z)$ with an FRW solution. We have simply compared the shape of 
the $d_L(z)$ for the swiss-cheese model with the one of a $\Lambda$CDM model.

We intend now to fit a phenomenological FRW model to our swiss-cheese model. The
FRW model we  have in mind is a spatially flat model with a matter component
with present fraction of the energy density $\Omega_M=0.25$, and with a
phenomenological dark-energy component with present fraction of the energy
density $\Omega_\Lambda=0.75$.  We will assume that the dark-energy component
has an equation of state
\begin{equation}
\label{fame}
w(a)=w_{0}+ w_{a}\left(1-\frac{a}{a_0}\right)=w_0+w_a \; \frac{z}{1+z} .
\end{equation}
Thus, the total energy density in the phenomenological model evolves as
\begin{equation} 
\label{linda}
\frac{\rho^\textrm{FIT}}{\rho_{0}} = \Omega_M(1+z)^3 + \Omega_\Lambda
(1+z)^{3(1+w_{0}+w_{a})}\; \exp\left(-3w_{a}\frac{z}{1+z}\right) .
\end{equation}
We will refer to this model as the {\it phenomenological model} throughout 
this paper.

Our swiss-cheese model is a lattice of holes as sketched in Fig.\
\ref{schizzo}:  the scale of the inhomogeneities is therefore simply the size of
a hole.  We are interested in understanding how the equation of state of the
``dark energy'' in the phenomenological model changes with respect to $r_{h}$,
and in  particular, why. Of course, in the limit $r_{h} \rightarrow 0$, we
expect to find  $w=0$, that is, the underlying EdS model out of which the cheese
is constructed.

\begin{figure}
\begin{center}
\includegraphics[width=13 cm]{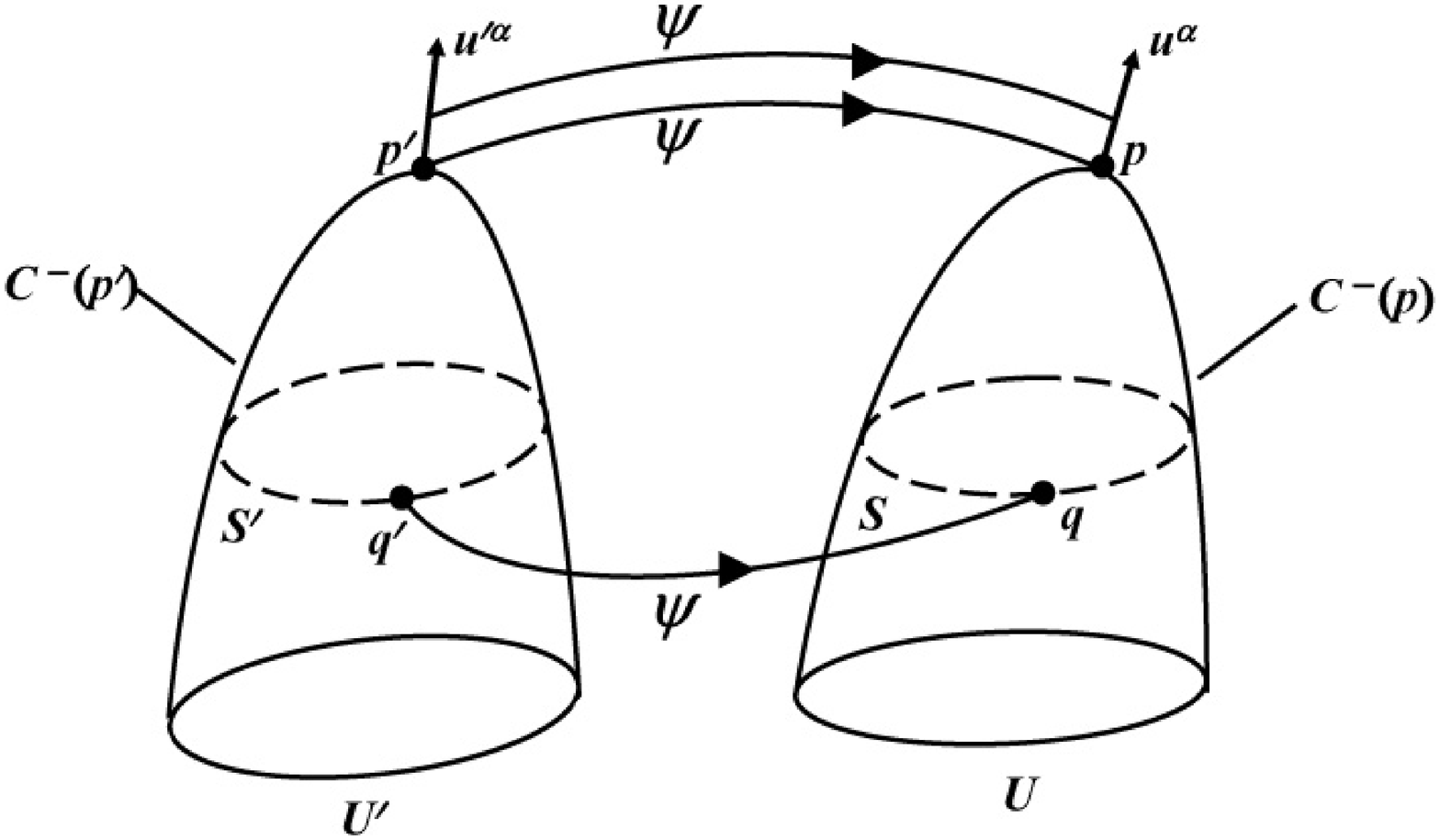}
\caption{In the null data best fitting, one successively chooses maps  from the
real cosmological model $U$ to the FRW model $U'$ of the null  cone vertex
$p'$, the matter 4-velocity at $p'$, a two-sphere $S'$ on the  null cone of
$p'$ and a point $q'$ on the 2-sphere. This establishes the  correspondence
$\psi$ of points on the past null cone of $p'$, $C^-(p^\prime)$, to the past 
null cone of $p$, $C^-(p^\prime)$, and then compares initial data at $q'$ and
at $q$. From  Figure 2 of \cite{ellis-f}.}
\label{map}
\end{center}
\end{figure}

The procedure developed by Ref.\ \cite{ellis-f} is summarized by Fig.\
\ref{map}.  We refer the reader to that reference for a more thorough analysis
and to Ref.\ \cite{Celerier:2007tp} and references therein for recent
developments. We will focus now in using our swiss-cheese model as (toy)
cosmological model.

\subsection{Choice of vertex points}

We start choosing the two observers to be compared. In the homogeneous FRW
model every observer is the same thanks to spatial  homogeneity. We choose an
observer in the cheese as the corresponding observer in our swiss-cheese model,
in particular the one shown in Fig.\ \ref{schizzo}.

Our model allows us to choose also the time of observation, which, in general,
is a final product of the comparison. We now explain why.

The FRW model we will obtain from the fit will evolve differently from the 
swiss cheese: the latter evolves as an EdS model, while the former will 
evolve as a quintessence-like model. They are really different models. 
They will agree only along the light cone, that is, on our observations.

Now, for consistency, when we make local measurements\footnote{Conceptually, 
it could not be possible with a realistic universe model to make 
local measurements that could be directly compared to the smoothed FRW 
model. We are allowed to do this thanks to our particular swiss-cheese 
model in which the cheese well represents the average properties of the 
model.} the two models have to give us the same answer: local
measurements indeed can be seen as averaging measurements with a small enough 
scale of averaging, and the two models agree along the past light cone. 

Therefore, we choose the time in order that the two observers measure  the same
local density. This feature is already inherent in Eq.\ (\ref{linda}):  the
phenomenological model and the swiss-cheese model evolve in order to have the
same local density, and therefore the same Hubble parameter, at the present
time.

\subsection{Fitting the 4-velocity}

The next step is to fit the four-velocities of the observers.  In the FRW model
we will choose a comoving observer, the only one who experiences an isotropic
CMB. In the swiss-cheese model, we will choose, for the same reason, a 
cheese-comoving observer. Again, our swiss-cheese model considerably simplifies
our work.

\subsection{Choice of comparison points on the null cones} \label{compa}

Now that the past null cones are uniquely determined, we have to choose a 
measure of distance to compare points along each null cone.

First, let us point out that instead of the entire two-sphere along the null
cone, we will examine, only a point on it. This is because of the simplified
set-up of our swiss-cheese model in which  the observer is observing only in
two opposite directions, as  illustrated in Fig.\ \ref{schizzo}. This means
that we can skip the step consisting in averaging our observable quantities
over the surface of constant redshift,  which generally is necessary in order
to be able to compare an  inhomogeneous model with the FRW model
\cite{ellis-f}.

Coming back to the main issue of this section,  we will use the observed
redshift  $z$ to compare points along the null cones. Generally, the
disadvantage of  using it is that it does not directly represent distances
along the null  cone. Rather, the observed value $z$ is related to the
cosmological redshift  $z_{C}$ by the relation:
\begin{equation}
1+z=(1+z_{O})(1+z_{C})(1+z_{S})
\end{equation}
where $z_{O}$ is the redshift due to the peculiar velocity of the observer 
$O$ and $z_{S}$ that due to the peculiar velocity of the source. The latter, 
in particular, is a problem because local observations cannot distinguish 
$z_{S}$ from $z_{C}$. 

However, our set up again simplifies this task. The observers chosen are, 
indeed, both comoving (in the swiss-cheese model because the observer is in the
cheese, and in the phenomenological model by construction), and therefore
$z_{O}=0$.  Regarding the sources,  we know exactly their behavior because we
have a model to work with.

The sources are also comoving; however, there are structure-formation  effects
that should be disentangled from the average evolution. For this  reason we
will perform averages between points in the cheese (the  meaning of this will
be clear in the next section) in order to  smooth out these structure-formation
effects.

\subsection{Fitting the null data} \label{fitting}

\begin{figure}
\begin{center}
\includegraphics[width=16.2 cm]{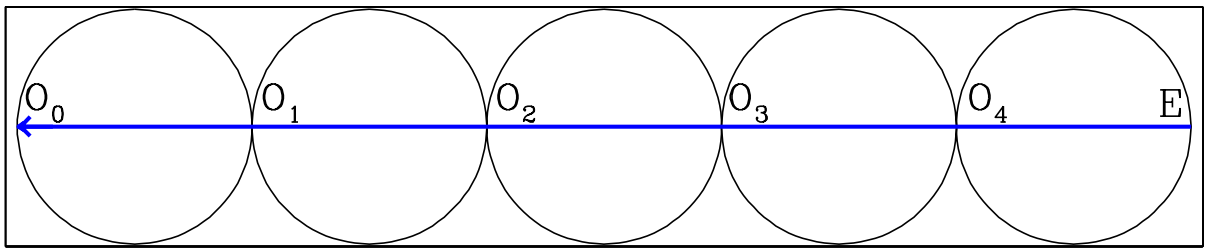}
\caption{An illustration of the points chosen for the averaging procedure.}
\label{shime}
\end{center}
\end{figure}

Now we are ready to set up the fitting of our swiss-cheese model. Ref.\
\cite{Hellaby:1988zz} studied the approach based on volume averaging outlined
in Ref.\ \cite{ellis-f}. This approach, however, is appropriate for studies 
concerning global dynamics, as in Refs.\ \cite{carfora, buchert_new}. As
stressed previously, here we are instead interested in averages  {\it directly}
related to observational quantities, and we constructed our  model following
this idea: it is a model that is exactly solvable and  ``realistic'' (even if
still toy) at the price of no interesting volume-averaged dynamics.

Therefore, we will follow a slightly different approach from the ones outlined
in Ref.\ \cite{ellis-f}:  we are going to fit averages along the light cone.
This method will be intermediate between the fitting approach and the
averaging approach.

We will focus on the expansion scalar and the density. We will see that these
two quantities behave differently under averaging. We denote by
$Q^\textrm{SC}(r, t)$ a quantity in the swiss-cheese model we want to average.
We denote by $Q^\textrm{FIT}(t)$ the corresponding quantity we want to fit to
the average of $Q^\textrm{SC}(r, t)$. Note that $Q^\textrm{FIT}(t)$ does not
depend on $r$ because the phenomenological model we will employ to describe the
swiss-cheese model is homogeneous.

Again, the fit model is a phenomenological homogeneous model (just refereed to
as the phenomenological model).  It need not be the model of the cheese.

The procedure is as follows. First we will average $Q^\textrm{SC}(r, t)$ for a
photon that starts from the emission point $E$ of the five-hole chain and
arrives at the locations of observers $O_{i}$ of Fig.\ \ref{shime}. We have
chosen those points because they well represent the average dynamics of the
model. Indeed, these points are not affected by structure evolution because they
are in the cheese.  Then, we will compare this result with the average of
$Q^\textrm{FIT}(t)$ for  the phenomenological and homogeneous source with
density given by Eq.\ (\ref{linda}) with an equation of state $w$ given by Eq.\
(\ref{fame}). 

The two quantities to be compared are therefore:
\begin{eqnarray}
\langle Q^\textrm{SC} \rangle_{\overline{\scriptscriptstyle EO}_{i}}
& = & \left[  \int_{E}^{O_{i}} dr \; Y' / W \right]^{-1} 
\int_{E}^{O_{i}} dr \ Q^\textrm{SC}(r,t(r)) \; Y'(r,t(r)) / W(r)  \nonumber \\
\langle Q^\textrm{FIT} \rangle_{\overline{\scriptscriptstyle EO}_{i}} 
\label{ququ1}
& = & \left[  \int_{E}^{O_{i}} dr \;  a_{\scriptscriptstyle FIT} \right]^{-1}
\int_{E}^{O_{i}} dr \ Q^\textrm{FIT}(t_{\scriptscriptstyle FIT}(r))\; 
a_{\scriptscriptstyle FIT}(t_{\scriptscriptstyle FIT}(r))  ,
\end{eqnarray}
where $t(r)$ and $t_{\scriptscriptstyle FIT}(r)$ are the photon  geodesics in
the swiss-cheese model and in the phenomenological one, respectively. The
functions $t_{\scriptscriptstyle FIT}(r)$, $a_{\scriptscriptstyle FIT}$  and
other  quantities we will need are obtained solving the Friedman  equations
with a source described by Eq.\ (\ref{linda}) with no curvature. The points
$O_{i}$ in the swiss-cheese model of Fig.\ \ref{shime}  are associated to
points in the phenomenological model with the same redshift, as discussed in
Sec.\ \ref{compa}.

We will then find the $w$ that gives the best fit between 
$\langle Q^\textrm{FIT}\rangle$ and $\langle Q^\textrm{SC}\rangle$, that 
is, the choice that minimizes:
\begin{equation} \label{ququ2}
\sum_{i} \left ( \langle Q^\textrm{FIT} 
\rangle_{\overline{\scriptscriptstyle EO}_{i}}-
\langle Q^\textrm{SC} 
\rangle_{\overline{\scriptscriptstyle EO}_{i}} \right )^{2}  .
\end{equation}
Of course, in the absence of inhomogeneities, this method would give $w=0$.

Let us summarize the approach:
\begin{itemize}

\item We choose a phenomenological quintessence-like model that, at the
present time, has the same density and Hubble parameter as the EdS-cheese
model. 

\item We make this phenomenological model and the swiss-cheese model
correspond along the light cone via light-cone averages of $Q$.

\item We can substitute the swiss-cheese model with the phenomenological 
model as far as the averaged quantity $Q$ is concerned.

\end{itemize} 

The ultimate question is if it is observationally meaningful to consider $Q$,
as opposed to the other choice of domain averaging at constant time, which is
not directly  related to observations. We will come back to this issue after
having obtained the results.

\subsubsection{Averaged expansion}

The first quantity in which we are interested is the expansion rate. To average
the expansion rate we will follow the formalism developed in Sec.\
\ref{histories}. We will therefore apply Eqs.\ (\ref{ququ1}-\ref{ququ2}) to 
$Q^\textrm{SC}=H_{r}\equiv\dot{Y}' / Y'$, where we remember that $H_r$ is the
radial expansion rate. The corresponding  quantity in the phenomenological
model is $Q^\textrm{FIT}=\dot{a}_{\scriptscriptstyle 
FIT}/a_{\scriptscriptstyle FIT}$.

\begin{figure}
\begin{center}
\includegraphics[width=13 cm]{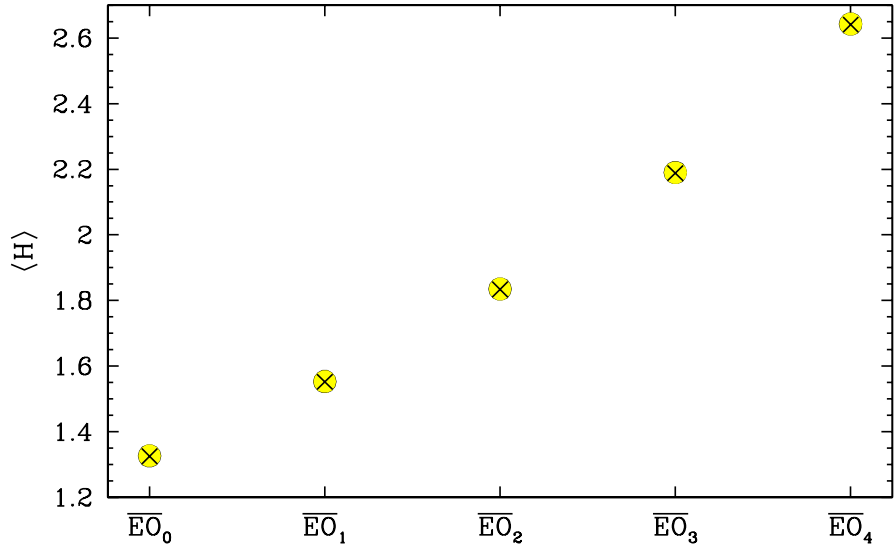}
\caption{Average expansion rate. The yellow points are $\langle H^\textrm{SC} 
\rangle_{\overline{\scriptscriptstyle EO}_{i}}$
 while the crosses are $\langle H^\textrm{FIT} 
\rangle_{\overline{\scriptscriptstyle EO}_{i}}$.
 $\overline{EO}_{i}$ means that the average was performed from $E$
and $O_{i}$ with respect to Fig.\ \ref{shime}. The best fit is 
found for
$w \simeq0$, that is, the phenomenological model is the cheese-FRW solution 
itself as far as the expansion rate is concerned.}
\label{hfit}
\end{center}
\end{figure}

For the same reason there is good compensation in redshift  effects (see Sec.\
\ref{histories}), we expect $\langle H_{r} \rangle$ to behave very similarly to
the FRW cheese solution. Indeed, as one can see in Fig.\ \ref{hfit}, the best
fit of the  swiss-cheese model is given by an phenomenological source with
$w\simeq0$, that  is, the phenomenological model is the cheese-FRW solution
itself as far as the  expansion rate is concerned.

\subsubsection{Averaged density}

\begin{figure}
\begin{center}
\includegraphics[width=13 cm]{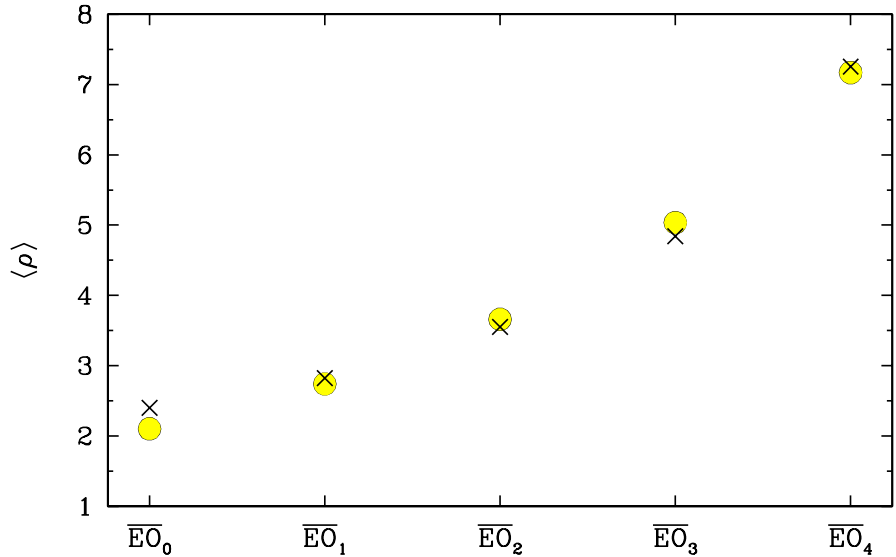}
\caption{Average density in $\rho_{C 0}$ units. The  yellow points
are  $\langle \rho^\textrm{SC} \rangle_{\overline{\scriptscriptstyle EO}_{i}}$
while the crosses are $\langle \rho^\textrm{FIT} 
\rangle_{\overline{\scriptscriptstyle EO}_{i}}$. $\overline{EO}_{i}$ means that
the average was performed from $E$ and $O_{i}$ with respect to Fig.\
\ref{shime}. The parametrization of  $\rho^\textrm{FIT}$ is from Eq.\
(\ref{linda}). The best fit is found for $w_{0}=-1.95$ and $w_{a}=4.28$.}
\label{qfit}
\end{center}
\end{figure}

The situation for the density is very different. The photon is spending more and
more time in the (large) voids than in the (thin) high density structures. We
apply Eqs.\ (\ref{ququ1}-\ref{ququ2}) to $Q^\textrm{SC}=\rho^\textrm{SC}$. The
corresponding quantity in the phenomenological model is 
$Q^\textrm{FIT}=\rho^\textrm{FIT}$ where $\rho^\textrm{FIT}$ is given by Eq.\
(\ref{linda}). The results are illustrated in Fig.\ \ref{qfit}: the best fit is
for  $w_{0}=-1.95$ and $w_{a}=4.28$.

As we will see in Sec.\ \ref{dressing}, we can achieve a better fit to the
concordance model with smaller holes than the ones of $350$ Mpc considered
here. We anticipate that for a holes of radius $r_{h}=250$ Mpc, we have
$w_{0}=-1.03$ and $w_{a}=2.19$.

We see, therefore, that this swiss-cheese model could be interpreted, in the 
FRW hypothesis, as a homogeneous model that is initially dominated by matter
and subsequently by dark energy: this is what the concordance model suggests.
We stress that this holds only for the light-cone averages of the density.

\section{Discussion} \label{disco}

\subsection{Explanation}

\begin{figure}
\begin{center}
\includegraphics[width=12 cm]{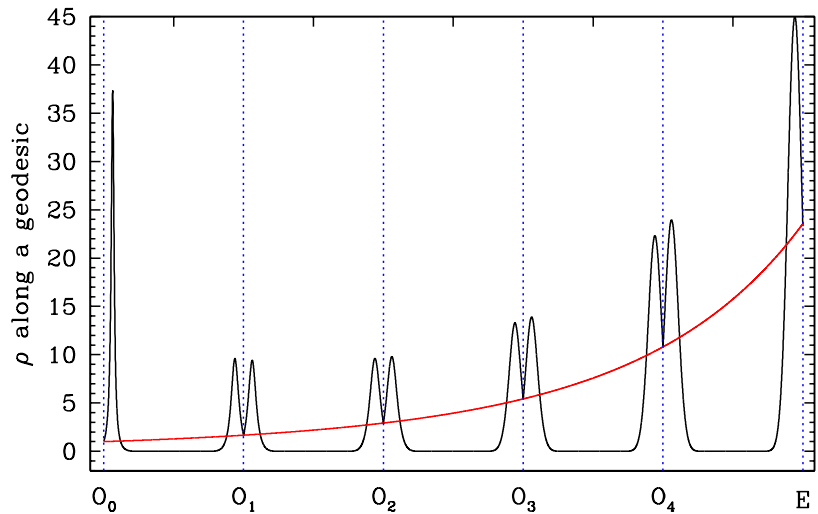}
\caption{Density along the light-cone for the swiss-cheese model (the spiky
curve) and the EdS model of the cheese (the regular curve). The labeling of 
the $x$-axis is the same one of Fig.\ \ref{shime}.}
\label{photorho}
\end{center}
\end{figure}

Let us first explore the basis for what we found. In  Fig.\ \ref{photorho} we
show the density along the light cone for both the swiss-cheese model and the
EdS model for the cheese. It is clear that the  photon is spending more and more
time in the (large) voids than in the (thin) high density structures.

\begin{figure}
\begin{center}
\includegraphics[width=11 cm]{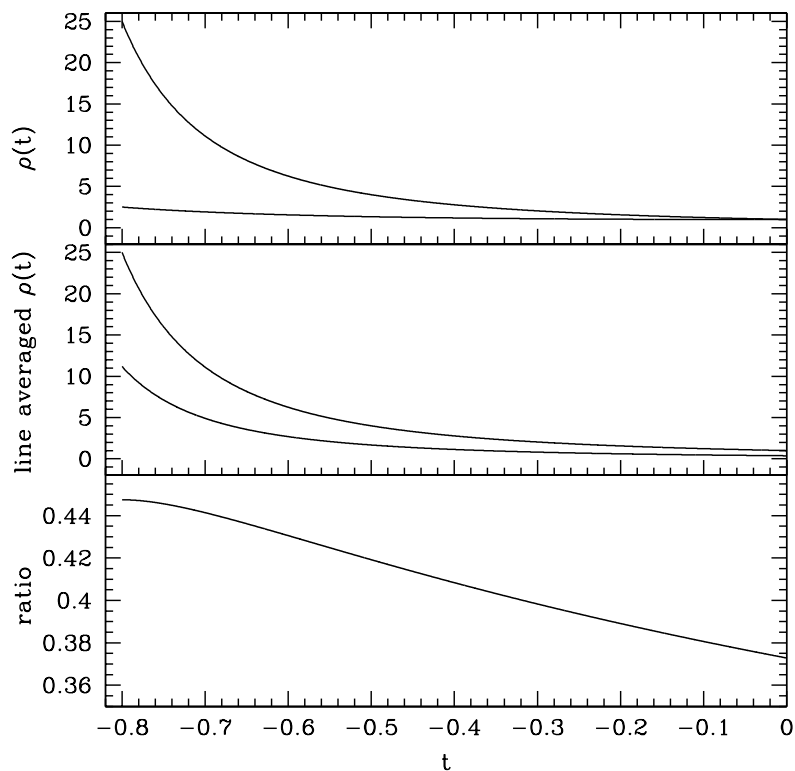}
\caption{At the top is the evolution of the energy density for the  Eds cheese
model (higher curve) and for the phenomenological model with  $w_{0}=-1.95$ and
$w_{a}=4.28$. In the middle is the constant-time line averaged density as a
function of  time for the swiss-cheese model (lower curve) and the cheese-EdS
model  (higher curve).  At the bottom is their ratio of the last two quantities
as  a function of time.}
\label{opacity}
\end{center}
\end{figure}

To better show this, we plotted in Fig.\ \ref{opacity} the constant-time, 
line-averaged density as a function of time. The formula used for 
the swiss-cheese model is
\begin{equation} \label{opacityf}
\int_{0}^{r_{h}} dr \ \rho(r,t) \; Y'(r,t) / W(r)  \left /  
\int_{0}^{r_{h}}dr \; Y' / W \right.  ,
\end{equation}
while for the cheese, because of homogeneity we can just use $\rho(t)$ of the
EdS model. As one can see, the photon is encountering less matter in the
swiss-cheese model than in the EdS cheese model. Moreover, this becomes
increasingly true with the formation of high-density regions as illustrated in 
Fig.\ \ref{opacity} by the evolution  of the ratio of the previously calculated
average density: it  decreases by $17 \%$ from the starting to the ending time. 

The calculation of Eq.\ (\ref{opacityf}) is actually, except for some factors 
like the cross-section, the opacity of the swiss-cheese model. Therefore, a
photon propagating through the swiss-cheese model has a different average
absorption history; that is, the observer looking through the cheese will
measure  a different flux compared to the case with only cheese and no holes.
For the moment, in order to explore the physics, let us make the approximation
that during the entire evolution of the universe, the matter is transparent to
photons.

{}From the plots just shown we can now understand the reason for the best fit 
values of $w_{0}=-1.95$ and $w_{a}=4.28$ found in the case of holes of
$r_{h}=350$ Mpc. We are using a homogeneous phenomenological model, which has at
the present time the  density of the cheese (see Fig.\ \ref{opacity}). We want
to use it to fit the line-averaged density of the swiss cheese, which is lower
than the (volume) averaged one. Therefore, going backwards from the present
time, the phenomenological model must keep its density low, that is, to have a
small $w$. At some point, however, the density has to start to increase,
otherwise it will not match the line-averaged value that keeps increasing:
therefore  $w$ has to increase toward $0$. It is very interesting that this
simple mechanism mimics the behavior of the concordance-model equation of state.
We stress that this simple mechanism works thanks to the set-up and fitting
procedure we have chosen; that is, the fact that we matched the cheese-EdS
solution at the border of the hole, the position of the observer, and the
observer looking through the holes. Moreover, we did not tune the model to
achieve a best matching with the concordance model. The results shown are indeed
quite natural.

\subsection{Beyond spherical symmetry} \label{cosca}

In this work we are interpreting the swiss-cheese model from the point of  view
of light-cone averages. In Ref.\ \cite{marra-sc} we have instead focused on the 
luminosity-distance--relation (see Fig.\ \ref{5incheese}).

\begin{figure}[h!]
\begin{center}
\includegraphics[width=16 cm]{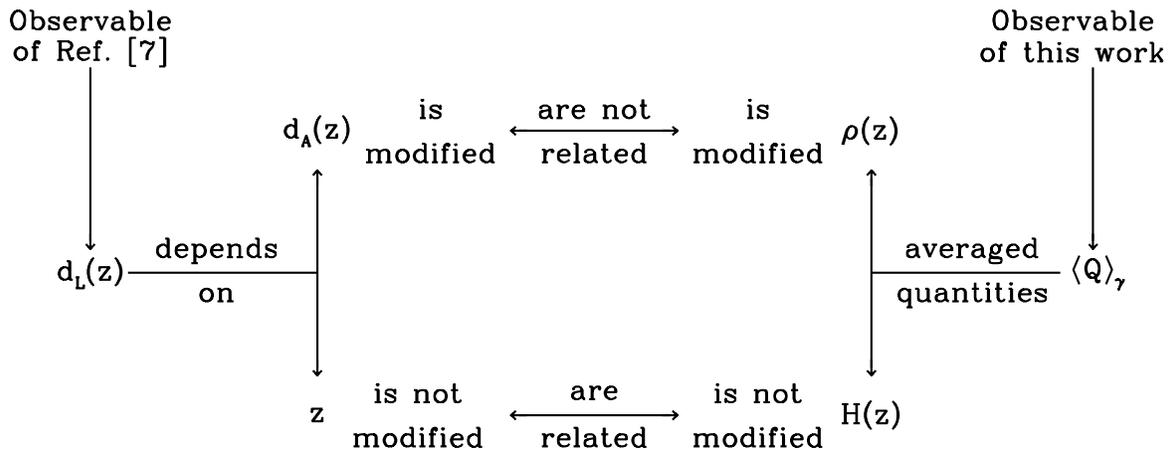}
\caption{Flow chart regarding relationships between the results obtained in
Ref.\ \cite{marra-sc} and this work   See Sec.\ \ref{cosca} for a discussion.}
\label{schema}
\end{center}
\end{figure}

We have summarized the relationships between the results obtained in Ref.\
\cite{marra-sc} and this work in the flow chart of  Fig.\ \ref{schema}.

Regarding $d_{L}(z)$, we found no important effects from a change in the
redshift:  the effects on $d_L(z)$ all came from $d_{A}$ driven by the evolution
of the inhomogeneities.

Regarding light-cone averages, we found no important effects with respect 
to the expansion: this negative result is due to the compensation in 
redshift discussed in Sec.\ \ref{histories} and it is the same reason 
why we did not find redshift effects with $d_{L}(z)$. This is the main 
limitation of our model and it is due ultimately to the spherical
symmetry of the model as explained in Sec.\ \ref{histories}.

We found important effects with respect to the density: however this is 
not due to the effects driving the change in $d_{A}$. The latter is due to 
structure evolution while the former to the presence of voids, so the 
two causes are not directly connected. Indeed, it is possible to turn off 
the latter and not the former.

We can therefore make the point that the expansion is not affected by  the
inhomogeneities because of the compensation due to the spherical symmetry. The
density, on the other hand, is not affected by the spherical symmetry, so there
are no compensations, and the photon systematically sees more and more voids
than  structures. We can therefore argue that the average of the density is more
relevant  than the average of the expansion because it is less sensitive to the
assumption of spherical symmetry,  which is one of the limitations of this
model.

The next step is to define a Hubble parameter from this average density:  $H^{2}
\propto \langle \rho \rangle_{\gamma}$. In this way we are moving  from a
swiss cheese made of spherically symmetric holes to a swiss cheese  without
exact spherical symmetry. The correspondence is through the light-cone averaged
density which,  from this point of view, can be seen as a tool in performing
this step.

Summarizing again:

\begin{itemize}

\item We started with a swiss-cheese model containing only spherically
symmetric  holes. A photon, during its journey through the swiss cheese,
undergoes a  redshift that is not affected by inhomogeneities. However the
photon is  spending more and more time in the voids than in the structures. The
lack of an effect is due to spherical symmetry. We focus on this because a
photon spending most of its time in voids should have a different redshift
history than a photon propagating in a homogeneous background.

\item Since the density is a quantity that is not particularly sensitive to
spherical symmetry, we try to resolve the mismatch  by focusing on the 
density alone and getting from it the expansion (and therefore the redshift 
history).

\item We ended up with a swiss-cheese model with holes that effectively are not
spherically symmetric. In this model there is an effect on the redshift history
of a photon due to the voids.

\item In practice this means that we will use the phenomenological best-fit
model  found, that is, we will use a model that behaves similarly to the
concordance  model.

\end{itemize}

\subsection{Motivations}

Let us go back to the discussion of Sec. \ref{fitting}, that is, if it is 
observationally meaningful to consider light-cone averages of $Q$  as the basis
for the correspondence. For example, domain averages at constant time are not
directly related  to observations.

Here, we are not claiming that light-cone averages are observationally 
relevant\footnote{However, a density light-cone average is an indicator 
of the opacity of the universe and therefore could be observationally 
relevant, as explained in the discussion around Fig.\ \ref{opacity}.}. 
Rather, we are using light-cone averages as tools to understand the model 
at hand. The approach has been explained in the previous section.

\section{Renormalization of the matter equation of state} \label{dressing}

\begin{figure}
\begin{center}
\includegraphics[width=14cm]{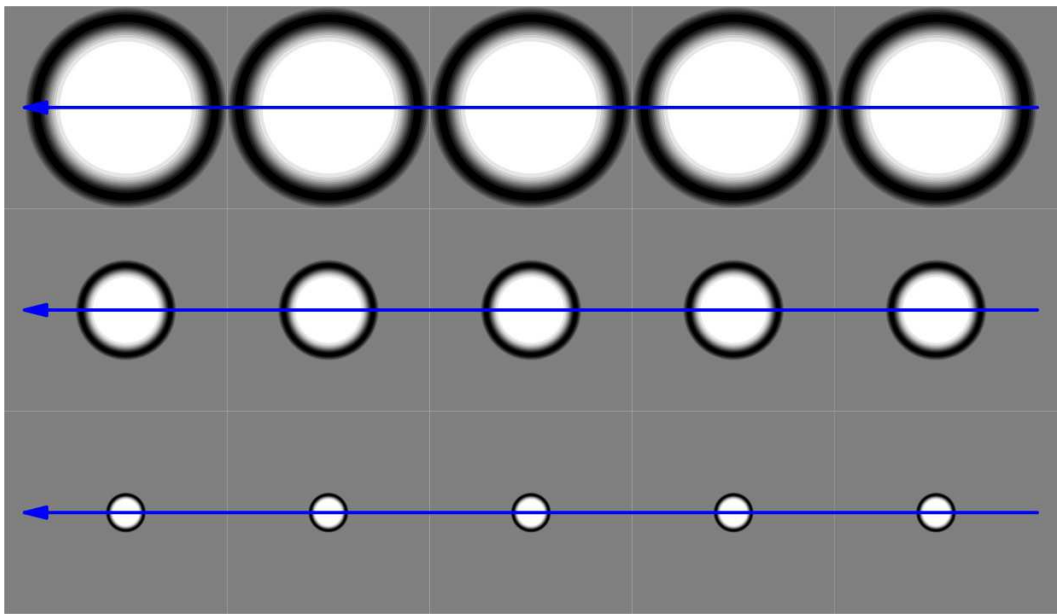}
\caption{Sketch of how the size of the inhomogeneity is changed in our  model.
The shading mimics  the initial density profile: darker shading implies larger
denser. The  uniform  gray is the FRW cheese. The photons pass through the holes
as shown by the  arrows and are revealed by the observer whose comoving position
in the  cheese does not change.  The size of the holes correspond to $n=0,$
$2$, $5$ of Eq.\ (\ref{scala}).}
\label{sdressing}
\end{center}
\end{figure}

In this section we will study how the parameters of the phenomenological  model
depend on the size of the inhomogeneities, that is, on the size  of the hole. We
sketched in Fig.\ \ref{sdressing} our set-up: we keep the comoving position of
the centers of the holes fixed. The observer is located in  the same piece of
cheese.

We changed the radius of the hole according to
\begin{equation} \label{scala}
r_h(n) = \frac{r_h}{1.4^{n}} ,
\end{equation}
where $r_{h}$ is the radius we have been using till now, the one that results in
the holes touching. The choice of the $1.4$ in the scaling is only  for
convenience. We let $n$ run from $0$ to $7$.

In this analysis we will use instead of the energy density in Eq.\
(\ref{linda}), an  energy density in which only one effective source appears,
and the effective source evolves as
\begin{equation} 
\label{linda2}
\frac{\rho^\textrm{FIT}}{\rho_{0}} =
(1+z)^{3(1+w^{R}_{0}+w^{R}_{a})}\; \exp\left(-3w^{R}_{a}\frac{z}{1+z}\right)
\qquad \mbox{with} \qquad
w^{R}(z)=w^{R}_0+w^{R}_a \; \frac{z}{1+z} .
\end{equation}

We put $R$ as a superscript on the equation of state in order to differentiate
the parametrization of Eq.\ (\ref{linda2}), which we are now using to study the
renormalization, from the parametrization of Eqs.\ (\ref{fame}-\ref{linda}),
which we used to compare the phenomenological model to the concordance model.
We are not disentangling different sources in Eq.\ (\ref{linda2}) because we
are interested in the renormalization of the matter equation of state of the
cheese,  that is, on the dependence of $w^{R}$ upon the size of the hole. To
this purpose we need only one source in order to keep track of the changes.

\begin{figure}
\begin{center}
\includegraphics[width=14.5 cm]{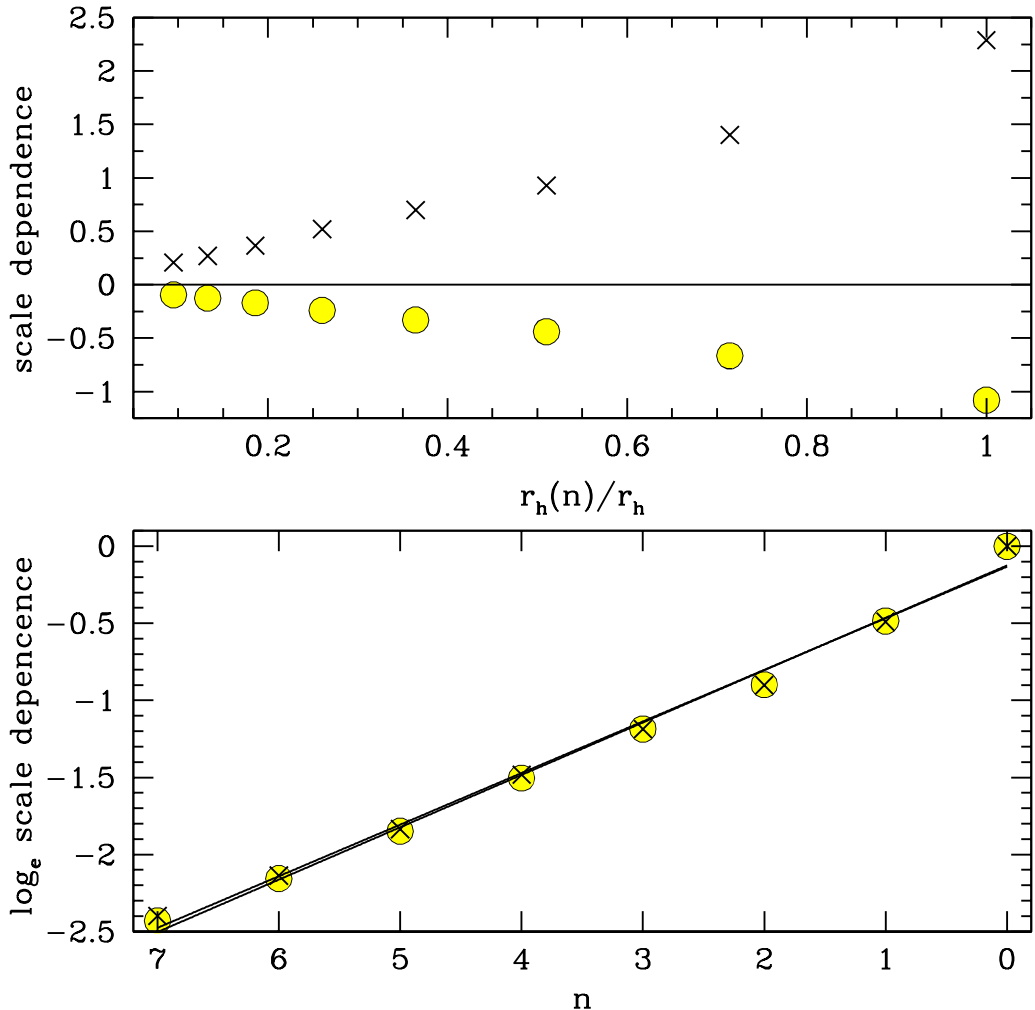}
\caption{At the top, dependence of $w^{R}_{0}$ (lower points denoted by circles)
and $w^{R}_{a}$  (upper points denoted by $\times$) with respect of the size of
the hole. At the bottom, fit as explained in the text. Recall that $r_{h}$ is
today $350$ Mpc.}
\label{dress}
\end{center}
\end{figure}

As one can see from Fig.\ \ref{dress},  we have verified that $w^{R}=0$ for
$r_{h} \rightarrow 0$, \textit{i.e.,} we recover the EdS model as the best-fit
phenomenological model.

We are interested to see if the equation of state exhibits a  power-law behavior
and, therefore, we use  the following functions to fit $w^{R}_{0}$ and
$w^{R}_{a}$:
\begin{eqnarray}
\frac{w^{R}_{0}(n)}{w^{R}_{0}(0)} & = & q_{0} \left ( \frac{r_{h}(n)}{r_{h}(0)} 
\right )^{p_{0}} \nonumber \\
\frac{w^{R}_{a}(n)}{w^{R}_{a}(0)} & = & q_{a} \left ( \frac{r_{h}(n)}{r_{h}(0)} 
\right )^{p_{a}} .
\end{eqnarray}
We performed a fit with respect to the logarithm of the above  quantities, the
result is shown in Fig.\ \ref{dress}. We found:
\begin{eqnarray}
p_{0} & = & p_{a} \simeq 1.00 \nonumber \\
q_{0} & = & q_{a} \simeq 0.88  .
\end{eqnarray}

Summarizing, we found three important facts.
\begin{itemize}

\item The parameters of the equation of state as a function of the 
size of the hole exhibit a power-law behavior.

\item The power-laws of $w^{R}_{0}$ and $w^{R}_{a}$ have the same scaling
exponent.  This is actually a check: once a physical quantity exhibits a
power-law  behavior, we expect that all its parameters share the same scaling
exponent.

\item The scale dependence is linear: the equation of state depends  linearly on
the length of holes the photon propagates through.  We stress that the
dependence we are talking about is not on the scale  of the universe, but on the
size of the holes.

\end{itemize}

We can finally ask which size of the holes will give us a phenomenological
model able to mimic the concordance model. We found that for $n=1$, that is for
a holes of radius $r_{h}=250$ Mpc, we have $w^R_0=1.4$ and $w^R_a=-0.665$,
which in terms of the energy density parametrization of Eq.\ (\ref{linda}),
corresponds to  $w_{0}=-1.03$  and $w_{a}=2.19$.

\section{Conclusions} \label{conclusions}

The aim of this investigation was to understand the role of large-scale
non-linear  cosmic inhomogeneities in the interpretation of observational data.
We focused on an exact (if toy) solution, based on the 
Lema\^{\i}tre-Tolman-Bondi (LTB) model. This solution has been studied 
extensively in the literature  \cite{alnes0607, notari-mansouri, alnes0602,
celerier, mansouri, flanagan, rasanen, tomita, chung, nambu}. It has been
shown  that it can be used to fit the observed $d_{L}(z)$  without the need of
dark energy (for example in Ref.\ \cite{alnes0602}).  To achieve this result,
however, it is necessary to place the observer at the center of a rather
large-scale underdensity. To overcome this  fine-tuning problem we built a
swiss-cheese model, placing the observer  in the cheese and having the observer
look through the holes in the swiss-cheese as pictured in Fig.\ \ref{schizzo}.

In Sec.\ \ref{model} we defined the model and  described its dynamics:  it is a
swiss-cheese model where the cheese is made of the usual FRW solution and the
holes are made of a LTB solution. The voids inside the holes  are expanding
faster than the cheese. We reported also the results for $d_{L}(z)$ obtained in 
Ref.\ \cite{marra-sc}, to which we refer the reader for a more thorough
analysis. We found that redshift effects are suppressed because of a
compensation effect due to spherical
symmetry. However, we found interesting effects in the calculation of the
angular distance: the evolution of the inhomogeneities bends the photon path
compared to the FRW case. Therefore, inhomogeneities will be able (at least
partly) to mimic the effects of dark energy.

After having analyzed the model from the observational point of view,  we set up
in Section \ref{fitti} the fitting problem in order to better  understand how
inhomogeneities renormalize the matter swiss-cheese model allowing us to eschew
a primary dark energy. We followed the scheme developed in Ref.\ \cite{ellis-f},
but modified in the way  to fit the phenomenological model to the swiss-cheese
one. We chose a method that is intermediate between the fitting approach  and
the averaging one: we fitted with respect to light-cone averages.

In particular, we focused on the expansion and the density. While the 
expansion behaved as in the FRW case because of the compensation  effect
mentioned above, we found that the density behaved differently thanks to  its
intensiveness to that compensation effect: a photon is spending more and more
time in the (large) voids than in the (thin) high density  structures. This
effect is not directly linked to the one giving us an  interesting $d_{A}$. The
best fit we found for holes of $r_{h}=250$ Mpc is $w_{0}=-1.03$ and
$w_{a}=2.19$; qualitatively similar to the concordance model.

The flow chart of Fig.\ \ref{schema} summarizes the results obtained. The
insensitivity to the compensation effect made us think that a swiss cheese made
of spherical symmetric holes and a swiss cheese without an exact spherical
symmetry  would share the same light-cone averaged density.  Knowing the
behavior of the density we are therefore able to know the one of the Hubble
parameter that will be the one of the FRW solution with an phenomenological
source characterized by the fit equation of state. In this way we can think to
go beyond the main limitation of this model, that is, the assumption of
spherical symmetry. From this point of view, the light-cone averaged density
can be seen as a  tool in performing this step.

Summarizing:
\begin{itemize}

\item We started with a swiss-cheese model based on spherically symmetric 
holes. A photon, during its journey through the swiss cheese, undergoes  a
redshift which is not affected by inhomogeneities. However the photon is 
spending more and more time in the voids than in the structures. The lack of an
effect is due to the the assumption of spherical symmetry. We focus on this
because a photon spending most of its time in voids should have a different
redshift history than a photon propagating in a homogeneous background.

\item Assuming that the density is a quantity that does not heavily depend  on
the assumption of spherical symmetry, we  tried to resolve the issue by
focusing on  the density alone and getting from it the expansion (and therefore
the  redshift history).

\item This resulted in a swiss-cheese model with holes that effectively are not
perfectly spherical. In this model the redshift history of a photon depends on
the time passed inside the voids. 

\item In practice this means that we will use the phenomenological best-fit
model  found, that is, we will use a model that behaves similarly to the
concordance  model.

\end{itemize}

Then, in Section \ref{dressing} we studied how the equation of state of  a
phenomenological model with only one effective source depends on the size of the inhomogeneity.
We found that $w^{R}_{0}$ and $w^{R}_{a}$ follow a power-law dependence with the same scaling 
exponent which is equal to unity. That is, the equation of state depends 
linearly on the distance the photon travels through voids.

We finally asked which size of the holes will give us a phenomenological model
able to mimic the concordance model.  We found that for $n=1$, that is for a
holes of radius $r_{h}=250$ Mpc, we have $w_{0}=-1.03$ and $w_{a}=2.19$.


\acknowledgments{It is a pleasure to thank Marie-No\"elle C\'el\'erier, George
Ellis and Antonio Riotto for useful discussions and suggestions.  V.M.
acknowledges support from ``Fondazione Ing. Aldo Gini'' and ``Fondazione Angelo
Della Riccia.'' }





\begin{thebibliography}{99}
\begin{frenchspacing}

\bibitem{kmr} 
E. W. Kolb, S. Matarrese, and A. Riotto, 
\textit{New J. Phys.} {\bf 8}, 322  (2006).

\bibitem{notari}
A. Notari,
Mod. Phys. Lett.  A {\bf 21}, 2997 (2006).

\bibitem{rasa}
S. Rasanen,
JCAP {\bf 0611}, 003 (2006).

\bibitem{buchert}
T. Buchert, J. Larena and J. M. Alimi,
Class. Quant. Grav.  {\bf 23}, 6379 (2006).

\bibitem{ellis}
G. F. R. Ellis,
Relativistic cosmology -- its nature, aims and problems.
In {\em General Relativity and Gravitation} (D. Reidel
Publishing Co., Dordrecht), ed. B. Bertotti, F. de Felice and 
A. Pascolini, pp. 215--288 (1984)

\bibitem{buchert_new} 
T. Buchert,
arXiv:0707.2153 [gr-qc].

\bibitem{marra-sc}
  V.~Marra, E.~W.~Kolb, S.~Matarrese and A.~Riotto,
  Phys.\ Rev.\  D {\bf 76}, 123004 (2007)
  [arXiv:0708.3622 [astro-ph]].

\bibitem{alnes0602} 
H. Alnes, M. Amarzguioui, and O. Gr{\o}n, 
\textit{Phys. Rev. D} {\bf 73}, 083519 (2006).

\bibitem{ellis-f}
G. F. R. Ellis and W. Stoeger, 
Class. Quantum Gravit. {\bf 4}, 1697 (1987).

\bibitem{lemaitre} 
A. G. Lema\^{\i}tre, 
\textit{Ann. Soc. Sci. Bruxelles} {\bf A53}, 51 (1933).

\bibitem{tolman} 
R. C. Tolman, 
\textit{Proc. Nat. Acad. Sci. USA} {\bf 20}, 169 (1934).

\bibitem{bondi} 
H. Bondi, 
\textit{Mon. Not. Roy. Astron. Soc.} {\bf 107}, 410 (1947).

\bibitem{einstein} 
A. Einstein and E. G. Straus, 
\textit{Rev. Mod. Phys.} {\bf 17}, 120 (1945).
  
\bibitem{Celerier:2007tp}
M. N. Celerier,
arXiv:0706.1029 [astro-ph].

\bibitem{Hellaby:1988zz}
C. Hellaby,
Gen. Rel. Grav. {\bf 20}, 1203 (1988).
  
\bibitem{carfora}
T. Buchert and M. Carfora,
Class. Quant. Grav. {\bf 19}, 6109 (2002).

\bibitem{alnes0607} 
H. Alnes and M. Amarzguioui, 
\textit{Phys. Rev. D} {\bf 74}, 103520 (2006).

\bibitem{notari-mansouri} 
T. Biswas, R. Mansouri and A. Notari,
arXiv:astro-ph/0606703.

\bibitem{celerier} 
M. N. C\'el\'erier, 
\textit{Astron. Astrophys.} {\bf 353}, 63 (2000).

\bibitem{mansouri} 
R. Mansouri,
astro-ph/0512605.

\bibitem{flanagan} 
R. A. Vanderveld, E. E. Flanagan, and I. Wasserman, 
\textit{Phys. Rev. D} {\bf 74},  023506 (2006).

\bibitem{rasanen} 
S. Rasanen, 
\textit{JCAP} {\bf 0411},  010 (2004).

\bibitem{tomita} 
K. Tomita, 
\textit{Prog. Theor. Phys.} {\bf 106},  929 (2001).

\bibitem{chung} 
D. J. H. Chung and A. E. Romano,
\textit{Phys. Rev. D} {\bf 74}, 103507 (2006).

\bibitem{nambu} 
T. Kai, H. Kozaki, K. Nakao, Y. Nambu and C. Yoto,
\textit{Prog. Theor. Phys.} {\bf 117},  229-240 (2007).


\end{frenchspacing}

\end{thebibliography}
\end{document}